\documentclass[aps,prd,preprintnumbers,groupedaddress,nofootinbib]{revtex4-1}

\usepackage{xspace}
\usepackage{graphicx}
\usepackage{amsmath}

\usepackage{url}

\DeclareRobustCommand{\Eq}[1]{Eq.~(\ref{#1})}

\DeclareRobustCommand{\Ref}[1]{Ref.~\cite{#1}}

\begin{document}

\preprint{MIT--CTP 4549}

\title{
Conformal Invariance of the Subleading Soft Theorem in Gauge Theory
}

\author{Andrew J. Larkoski}

\affiliation{Center for Theoretical Physics, Massachusetts Institute of Technology, Cambridge, MA 02139, USA}

\email{larkoski@mit.edu}

\begin{abstract}
In this note, I show that the recently proposed subleading soft factor in massless gauge theory uniquely follows from conformal symmetry of tree-level gauge theory amplitudes in four dimensions.
\end{abstract}

\maketitle

Recently, Cachazo and Strominger proposed a new soft theorem for gravity motivated by a conjecture of an enhanced symmetry of the quantum gravity S-matrix \cite{Cachazo:2014fwa}.  Shortly after their paper, it was pointed out that a similar soft theorem exists for gauge theory \cite{Casali:2014xpa}.\footnote{Other recent work includes \cite{Schwab:2014xua,Bern:2014oka,He:2014bga}.  Actually, both the subleading soft theorems in gauge theory and gravity have been known for some time.  The gauge theory soft theorem is called the Low-Burnett-Kroll theorem \cite{Low:1958sn,Burnett:1967km} and the gravity soft theorem was studied by Gross and Jackiw \cite{Gross:1968in,Jackiw:1968zza} and expressed in its modern form by White \cite{White:2011yy}.  I thank Duff Neill and Roman Jackiw for pointing out these references.}  Unlike gravity, massless tree-level gauge theory amplitudes in four dimensions are invariant under conformal transformations and this extra symmetry constrains the possible form of the subleading soft factor.  With the conformal invariance as a guide, I will show that the subleading soft theorem in gauge theory at tree-level is uniquely determined.

To determine the soft theorems, we consider a color-ordered and coupling-stripped $n$-point amplitude ${\cal A}_n$ in pure Yang-Mills gauge theory written in the spinor-helicity formalism \cite{Mangano:1990by,Dixon:1996wi}.  The amplitude can then be expressed as a function of the holomorphic $\lambda_i$ and anti-holomorphic $\tilde{\lambda}_i$ two-component spinors for particle $i$:
\begin{equation}
{\cal A}_n = \delta^{(4)}\left( \sum_i \lambda_i \tilde{\lambda}_i \right) A(1,\dotsc,n) \ ,
\end{equation}
where the $\delta$-function enforces momentum conservation and $A(1,\dotsc,n)$ is referred to as the stripped amplitude.  As explicitly shown in \Ref{Witten:2003nn} for MHV amplitudes, tree-level amplitudes in gauge theory are annihilated by the generators of the conformal group acting on the spinors $\lambda_i$ and $\tilde{\lambda}_i$.

To identify the soft behavior of the amplitude, we scale the momentum of a particle in the amplitude by a parameter $\epsilon$ and expand the amplitude in powers of $\epsilon$.  For a $+$ helicity particle $s$, this is most conveniently accomplished by a holomorphic scaling:
\begin{equation}
\lambda_s \to \epsilon \lambda_s\ , \qquad \tilde{\lambda}_s \to \tilde{\lambda}_s \ . 
\end{equation}
Then, the stripped amplitude has the form
\begin{equation}\label{eq:exp}
A(1,\dotsc,n,\{\epsilon \lambda_s,\tilde{\lambda}_s,+\}) = \left[\frac{1}{\epsilon^2}S^{(0)}(n,s,1) +\frac{1}{\epsilon}S^{(1)}(n,s,1)\right]A(1,\dotsc,n)+{\cal O}(\epsilon^0) \ ,
\end{equation}
where the soft factors are
\begin{equation}\label{eq:softfact0}
S^{(0)}(n,s,1) = \frac{\langle n 1 \rangle}{\langle ns\rangle \langle s 1\rangle} \ ,
\end{equation}
and 
\begin{equation}\label{eq:subsoft}
S^{(1)}(n,s,1) = \frac{\tilde{\lambda}_s^{\dot{a}}}{\langle s1 \rangle}\frac{\partial}{\partial \tilde{\lambda}_1^{\dot{a}}}+\frac{\tilde{\lambda}_s^{\dot{a}}}{\langle ns \rangle}\frac{\partial}{\partial \tilde{\lambda}_n^{\dot{a}}}\ .
\end{equation}
The spinor products are $\langle ij\rangle = \epsilon_{ab}\lambda_i^a\lambda_j^b$ and $[ ij] = \epsilon_{\dot{a}\dot{b}}\tilde{\lambda}_i^{\dot{a}}\tilde{\lambda}_j^{\dot{b}}$.  Because of color ordering, the soft factors only depend on the momenta of particles adjacent to the soft particles.  This property will be exploited throughout this note.

At this point, $\epsilon$ is an arbitrary expansion parameter that tracks the momentum of particle $s$, but is not assumed to be small.  Taking $\epsilon\to 1$ returns the full amplitude.  In particular, because the full amplitude is conformally invariant, then so is each term at a given order in $\epsilon$.  
Lorentz symmetry, momentum conservation and dilations are almost trivially satisfied on both terms because the stripped amplitude $A(1,\dotsc,n)$ is expressed as a function of the Lorentz covariant spinor products and has uniform mass dimension $4-n$.  Special conformal transformations, on the other hand, will provide non-trivial constraints on the soft factors.

The soft factors are defined in terms of the expansion of the stripped amplitude, \Eq{eq:exp}, but it is the full amplitude, which includes the momentum-conserving $\delta$-function that is invariant under conformal transformations.  So, to be able to use conformal invariance as a constraint on the soft factors, we must verify that conformal invariance provides a concrete constraint on the stripped amplitude alone, which can then be applied to the soft expansion.  This is what we turn to now.

In terms of the spinors, the special conformal generator $K_{a\dot{a}}$ is expressed as
\begin{equation}
K_{a\dot{a}} = \sum_i \frac{\partial^2}{\partial \lambda_i^a \partial \tilde{\lambda}_i^{\dot{a}}} \ ,
\end{equation}
where the sum runs over all particles in the amplitude.  The action of the special conformal generator on the full $n$-point amplitude ${\cal A}_{n}$ is
\begin{align}
K_{a\dot{a}} {\cal A}_{n} &= \left[ K_{a\dot{a}} \delta^{(4)}\left( \sum_i \lambda_i \tilde{\lambda}_i \right)\right] A(1,\dotsc,n) + \delta^{(4)}\left( \sum_i \lambda_i \tilde{\lambda}_i \right) K_{a\dot{a}}A(1,\dotsc,n) \nonumber \\
&\qquad+ \left[
\frac{\partial}{\partial P^{b\dot{a}}}\delta^{(4)}\left( \sum_i \lambda_i \tilde{\lambda}_i \right)
\right]\sum_i \lambda_i^{b}\frac{\partial}{\partial \lambda_i^a} A(1,\dotsc,n)\nonumber \\
&\qquad+ \left[
\frac{\partial}{\partial P^{a\dot{b}}}\delta^{(4)}\left( \sum_i \lambda_i \tilde{\lambda}_i \right)
\right]\sum_i \tilde{\lambda}_i^{\dot{b}}\frac{\partial}{\partial \tilde{\lambda}_i^{\dot{a}}} A(1,\dotsc,n) \ ,
\end{align}
where
\begin{equation}
P^{a\dot{a}} = \sum_i \lambda_i^{a}\tilde{\lambda}_i^{\dot{a}} \ .
\end{equation}
It was shown in \Ref{Witten:2003nn} that
\begin{equation}
K_{a\dot{a}} \delta^{(4)}\left( \sum_i \lambda_i \tilde{\lambda}_i \right) = (n-4)\frac{\partial}{\partial P^{a\dot{a}}}\delta^{(4)}\left( \sum_i \lambda_i \tilde{\lambda}_i \right) \ .
\end{equation}
Also, the terms with a single derivative on the momentum-conserving $\delta$-function can be simplified because the generators of Lorentz symmetry annihilate the stripped amplitude.  That is,
\begin{align}
&\frac{\partial}{\partial P^{b\dot{a}}}\delta^{(4)}\left( \sum_i \lambda_i \tilde{\lambda}_i \right)
\sum_i \lambda_i^{b}\frac{\partial}{\partial \lambda_i^a} A(1,\dotsc,n)+
\frac{\partial}{\partial P^{a\dot{b}}}\delta^{(4)}\left( \sum_i \lambda_i \tilde{\lambda}_i \right)
\sum_i \tilde{\lambda}_i^{\dot{b}}\frac{\partial}{\partial \tilde{\lambda}_i^{\dot{a}}} A(1,\dotsc,n)\nonumber \\
&=\frac{\partial}{\partial P^{b\dot{a}}}\delta^{(4)}\left( \sum_i \lambda_i \tilde{\lambda}_i \right)
\frac{1}{2}\delta_a^b\sum_i \lambda_i^{c}\frac{\partial}{\partial \lambda_i^c} A(1,\dotsc,n)\nonumber \\
&\qquad+\frac{\partial}{\partial P^{a\dot{b}}}\delta^{(4)}\left( \sum_i \lambda_i \tilde{\lambda}_i \right)
\frac{1}{2}\delta_{\dot{a}}^{\dot{b}}\sum_i \tilde{\lambda}_i^{\dot{c}}\frac{\partial}{\partial \tilde{\lambda}_i^{\dot{c}}} A(1,\dotsc,n)\nonumber\\
&=-(n-4)\frac{\partial}{\partial P^{a\dot{a}}}\delta^{(4)}\left( \sum_i \lambda_i \tilde{\lambda}_i \right) A(1,\dotsc,n) \ ,
\end{align}
where we have used the action of the dilation operator on the stripped amplitude to get the final line.  Combining these results, we find the action of the special conformal generator on the full amplitude is
\begin{equation}\label{eq:specconfdel}
K_{a\dot{a}} {\cal A}_n= \delta^{(4)}\left( \sum_i \lambda_i \tilde{\lambda}_i \right) K_{a\dot{a}}A(1,\dotsc,n) \ ,
\end{equation}
where all other terms explicitly cancel.  Therefore, for the full amplitude to be conformally invariant, it must be that $K_{a\dot{a}}A(1,\dotsc,n)=0$.  Importantly, at this point note that no approximations have been made nor any soft expansions performed.  

Using this result, we expand the action of the special conformal generator on the stripped amplitude in powers of $\epsilon$ subject to the constraint that
\begin{equation}
K_{a\dot{a}} A(1,\dotsc,n,s) = 0  \ ,
\end{equation}
which, by \Eq{eq:specconfdel}, is sufficient for enforcing invariance under special conformal transformations.
Note, that because $K_{a\dot{a}}$ depends on $\lambda_s$, we must also scale it appropriately.  That is, we will consider the scaled special conformal generator
\begin{equation}
K_{a\dot{a}}=\sum_{i=1}^{n}\frac{\partial^2}{\partial \lambda_i^a\partial\tilde{\lambda}_i^{\dot{a}}}+\frac{1}{\epsilon}\frac{\partial^2}{\partial \lambda_s^a\partial\tilde{\lambda}_s^{\dot{a}}} \ .
\end{equation}
Now, we can verify invariance under special conformal transformations order-by-order in $\epsilon$.  That is, we consider
\begin{align}\label{eq:expeq}
\left[
K_{a\dot{a}}A(1,\dotsc,n,\{\epsilon \lambda_s,\tilde{\lambda}_s,+\})
\right]_{\epsilon}&=
\left(
\sum_{i=1}^{n}\frac{\partial^2}{\partial \lambda_i^a\partial\tilde{\lambda}_i^{\dot{a}}}+\frac{1}{\epsilon}\frac{\partial^2}{\partial \lambda_s^a\partial\tilde{\lambda}_s^{\dot{a}}}
\right)\nonumber\\
&\quad\times\left(
\left[\frac{1}{\epsilon^2}S^{(0)}(n,s,1) +\frac{1}{\epsilon}S^{(1)}(n,s,1)\right]A(1,\dotsc,n)+{\cal O}(\epsilon^0)\right)
\ .
\end{align}
Order-by-order in $\epsilon$, the terms must vanish by conformal symmetry.  We will study the first few orders and will find that for consistency must demand that the subleading soft factor is precisely as defined in \Eq{eq:subsoft}.

At lowest order in $\epsilon$ from \Eq{eq:expeq}, we have
\begin{equation}\label{eq:leadingconf}
\left.K_{a\dot{a}}A(1,\dotsc,n,s)\right|_{\epsilon^{-3}}=\frac{1}{\epsilon^3}\frac{\partial^2}{\partial \lambda_s^a\partial\tilde{\lambda}_s^{\dot{a}}}  \left[S^{(0)}(n,s,1) A(1,\dotsc,n)\right] = 0\ , 
\end{equation}
because the soft factor is independent of $\tilde{\lambda}_s$ and the amplitude is fully independent of particle $s$.\footnote{Conversely, \Eq{eq:leadingconf} can be used to uniquely determine the leading soft factor.  Knowing that $S^{(0)}(n,s,1)$ is independent of one of $\lambda_s$ and $\tilde{\lambda}_s$ and using scaling, Lorentz and little group properties, the unique solution to \Eq{eq:leadingconf} is the familiar soft factor, \Eq{eq:softfact0}.}  At the next order, we have
\begin{align}
\left.K_{a\dot{a}}A(1,\dotsc,n,s)\right|_{\epsilon^{-2}}&=\frac{1}{\epsilon^2}\sum_{i=1}^{n}\frac{\partial^2}{\partial \lambda_i^a\partial\tilde{\lambda}_i^{\dot{a}}} \left[S^{(0)}(n,s,1) A(1,\dotsc,n)\right] \nonumber \\
&\qquad+\frac{1}{\epsilon^2}\frac{\partial^2}{\partial \lambda_s^a\partial\tilde{\lambda}_s^{\dot{a}}}  \left[S^{(1)}(n,s,1) A(1,\dotsc,n)\right] \nonumber \\
&= \frac{1}{\epsilon^2}\left[\frac{\partial}{\partial \lambda_n^a} S^{(0)}(n,s,1)\frac{\partial}{\partial\tilde{\lambda}_n^{\dot{a}}} +\frac{\partial}{\partial \lambda_1^a} S^{(0)}(n,s,1)\frac{\partial}{\partial\tilde{\lambda}_1^{\dot{a}}} \right]A(1,\dotsc,n)\nonumber \\
&\qquad+\frac{1}{\epsilon^2}\frac{\partial^2}{\partial \lambda_s^a\partial\tilde{\lambda}_s^{\dot{a}}}  S^{(1)}(n,s,1) A(1,\dotsc,n) \ ,
\end{align}
where the conformal invariance of the amplitude $A(1,\dotsc,n)$ has been used.  Evaluating the derivatives on the soft factor $S^{(0)}(n,s,1)$ we find
\begin{equation}\label{eq:soft0diff}
\frac{\partial}{\partial \lambda_n^a} S^{(0)}(n,s,1)\frac{\partial}{\partial\tilde{\lambda}_n^{\dot{a}}} +\frac{\partial}{\partial \lambda_1^a} S^{(0)}(n,s,1)\frac{\partial}{\partial\tilde{\lambda}_1^{\dot{a}}}= \frac{\lambda_{na}}{\langle ns \rangle^2}\frac{\partial}{\partial\tilde{\lambda}_n^{\dot{a}}}-\frac{\lambda_{1a}}{\langle s1 \rangle^2}\frac{\partial}{\partial\tilde{\lambda}_1^{\dot{a}}} \ .
\end{equation}

Being agnostic as to the form of $S^{(1)}$, we can determine it uniquely by demanding conformal invariance.  For the amplitude to be conformally invariant to this order in $\epsilon$, the derivatives on the subleading soft factor must be the opposite of \Eq{eq:soft0diff} and this highly constrains the possible form of the soft factor $S^{(1)}(n,s,1)$.  By the derivative structure in \Eq{eq:soft0diff}, for conformal invariance, the subleading soft factor must have the form
\begin{equation}
S^{(1)}(n,s,1) = {\cal F}(n,s,1)\frac{\partial}{\partial \tilde{\lambda}_n}+{\cal G}(n,s,1)\frac{\partial}{\partial \tilde{\lambda}_1} \ ,
\end{equation}
where spinor indices have been suppressed.  The functions ${\cal F}$ and ${\cal G}$ are constrained by mass dimension and the helicities of the particles $n,s,1$.  In particular, ${\cal F}$ must be independent of $\tilde{\lambda}_n$ by \Eq{eq:soft0diff}, and so for the soft factor to have zero net helicity for particle $n$ it must have the form
\begin{equation}
{\cal F}(n,s,1)=\frac{f(n,s,1)}{\langle ns\rangle} \ .
\end{equation}
For ${\cal F}$ to have the correct helicity of particle $s$ and mass dimension, it is therefore uniquely fixed to be
\begin{equation}
{\cal F}(n,s,1)=\frac{\tilde{\lambda}_s}{\langle ns\rangle} \ .
\end{equation}
This is precisely the correct form of the term in the subleading soft factor containing $n$, \Eq{eq:subsoft}.  Similar arguments constrain ${\cal G}$.

From the form of the subleading soft factor, conformal invariance can be verified explicitly.  The derivative on the soft factor $S^{(1)}(n,s,1)$ is
\begin{equation}
\frac{\partial^2}{\partial \lambda_s^a\partial\tilde{\lambda}_s^{\dot{a}}}  S^{(1)}(n,s,1) = -\frac{\lambda_{na}}{\langle ns \rangle^2}\frac{\partial}{\partial\tilde{\lambda}_n^{\dot{a}}}+\frac{\lambda_{1a}}{\langle s1 \rangle^2}\frac{\partial}{\partial\tilde{\lambda}_1^{\dot{a}}} \ .
\end{equation}
This is the opposite of \Eq{eq:soft0diff} and so
\begin{equation}
\left.K_{a\dot{a}}A(1,\dotsc,n,s)\right|_{\epsilon^{-2}}=0 \ ,
\end{equation}
proving that conformal invariance is preserved to this order in $\epsilon$.  One can continue to higher orders in $\epsilon$ and show that the tower of soft factors introduced in \Ref{He:2014bga} follow from enforcing conformal symmetry, also.

\begin{acknowledgments}
I thank Freddy Cachazo, Andy Strominger, Eduardo Casali, Anastasia Volovich, Song He, Congkao Wen, and especially Yu-tin Huang and Duff Neill for helpful discussions and correspondence.  This work is supported by the U.S. Department of Energy (DOE) under cooperative research agreement DE-FG02-05ER-41360. 
\end{acknowledgments}

\appendix

\bibliography{softgluons}

\end{document}